\documentclass[fleqn,usenatbib]{mnras}

\usepackage{newtxtext,newtxmath}

\usepackage[T1]{fontenc}
\usepackage{lscape}
\usepackage{multirow}
\usepackage{amsmath} 
\newcommand{\angstrom}{\textup{\AA}}
\DeclareRobustCommand{\VAN}[3]{#2}
\let\VANthebibliography\thebibliography
\def\thebibliography{\DeclareRobustCommand{\VAN}[3]{##3}\VANthebibliography}

\usepackage{graphicx}	
\usepackage{amsmath}	
\usepackage{siunitx}

\title[]{The origin of large emission line widths in massive galaxies at redshifts $z\sim 3-4$}

\author[]{M. Mart\'inez-Mar\'in
$^{1}$,\thanks{E-mail: mmartinezmarin@swin.edu.au}
K. Glazebrook$^{1}$,
T. Nanayakkara$^{1}$,
C. Jacobs$^{1}$,
I. Labb\'e$^{1}$,
G. G. Kacprzak$^{1}$, \newauthor 
C. Papovich $^{2}$
and C. Schreiber $^{3}$
\\
$^{1}$ Centre for Astrophysics and Supercomputing, Swinburne University of Technology, Hawthorn, VIC 3122, Australia \\
$^{2}$ Department of Physics and Astronomy, and George P. and Cynthia Woods Mitchell Institute for Fundamental Physics and Astronomy,\\ Texas AM University, College Station, TX 77843-4242, USA. \\
$^{3}$ IBEX Innovations, Sedgefield, Stockton-on-Tees, TS21 3FF, United Kingdom
}
\date{Accepted XXX. Received YYY; in original form ZZZ}

\pubyear{2023}

\begin{document}
\label{firstpage}
\pagerange{\pageref{firstpage}--\pageref{lastpage}}
\maketitle

\begin{abstract}

We present a sample of 22 massive galaxies with stellar masses $>10^{10} M_{\odot}$ at $3<z<4$ with deep H and K-band high resolution spectra (R=3500-3000) from Keck/MOSFIRE and VLT/KMOS near-infrared spectrographs. We find a large fraction have strong [OIII]5007 and H$\beta$ emission lines with large line widths ($\sigma$  100 -- 450 km/s). We measure the sizes of our galaxies  from Hubble Space Telescope images and consider the potential kinematic scaling relations of our sample; and rule out an explanation for these broad lines in terms of galaxy-wide kinematics. Based on consideration of the [OIII]5007 $/$ H$\beta$ flux ratios, their location in the Mass--Excitation diagram, and the derived bolometric luminosities, we conclude that Active Galactic Nuclei (AGN) and their Narrow Line Regions most likely give rise to this emission. At redshifts $3<z<4$, we find significantly high AGN fractions in massive galaxies, ranging from 60--70\% for the mass range $10<\log(M_{\star}/M_{\odot})<11$, with a lower limit 30\% for all galaxies within that redshift range when we apply our most stringent AGN criteria. We also find a considerably lower AGN fraction in massive quiescent galaxies, ranging from 20-30\%. These fractions of AGN point to the period between $3<z<4$ being a time of heightened activity for the development of supermassive black holes in the massive end of the galaxy population and provide evidence for their role in the emergence of the first massive quenched galaxies at this epoch.

\end{abstract}




\section{Introduction}

Massive galaxies in the early universe appear to assemble extremely rapidly. Spectroscopic detection of massive quiescent galaxies (QGs) of log$(M_{\star}/M_{\odot})>11$ at $3<z<4$ suggests that the stellar content was already in place in the first $1.5$ Gyr of the Universe's history
\citep{Marsan2017ApJ, Glazebrook2017, Tanaka_2019, Valentino2021, Forrest2020, esdaile2021, Forrest2022}.
The low-redshift stellar mass function requires that these compact galaxies both grow in size (about 3–5 times their effective radius and size-scaling relation) and quench quickly in order to evolve into the early-type galaxies observed in the local universe. Moreover, the number density of massive quiescent galaxies  at $z>3$ exceeds the predictions from galaxy
formation simulations by a factor of up to ten (\citealt[]{Schreiber2018}, hereafter S18; \citealt[]{Donnari2021, Lustig2022}).

The physical process that causes the rapid quenching and suppression of star formation in massive QGs is unclear. Both external and internal mechanisms can influence quenching. External phenomena such as ram-pressure stripping, tidal forces, and major or minor mergers can remove or heat the gas in group and cluster environments, affecting the gas reservoirs of galaxies. Internal processes like active galactic nuclei (AGN) can generate strong winds that can remove gas from the galaxy preventing star formation for a short period or permanently \citep{refId0, Feruglio, FischerrefId0, Komugi2023, Vayner2024} or may also produce positive feedback, as AGN outflows can trigger star formation by compressing cold, dense gas \citep{Magnelli, Silk, Zinn, Carniani2018, Shin2019}.

The efficiency of internal and external quenching mechanisms correlates with the galaxy stellar mass. Low-mass galaxies are more likely to be quenched by environmental effects, while in high-mass galaxies quenching is dominated by internal processes \citep{Peng2010, Baldry2006, Somerville2015}. The star formation activity of AGN hosts can change with redshift. At low redshifts, AGN are predominantly present in galaxies with suppressed star formation activity \citep{Ho, Salim2007}, while this trend weakens at high redshifts, as AGN predominantly reside in star-forming galaxies \citep{Cowley2016}.  

AGNs emit radiation across the entire electromagnetic spectrum, as the different emission processes depend on the geometry of the AGN and the line of sight \citep{Antonucci1993, Urry1995}. Therefore they can be detected through many observational techniques, such as broad and narrow optical emission lines \citep{Baldwin1981, Kewley_2001, Kauffmann2003}, X-ray emission \citep{Brandt_2015}, mid-infrared emission \citep{Lacy2004, Stern2005, Jarrett2011, Salyk_2019}, and radio emission \citep{Heckman_2014, Padovani2017}.

The x-ray and radio properties of typical QGs at $z>3$ are barely explored, mainly by individual detected sources that do not trace the entire AGN population and are biased towards high luminosity galaxies. Nevertheless, enhanced AGN activity is found in the QG population with high mass at $0<z<5$ \citep{Ito_2022}. \citet{Marsan2017ApJ}, from a sample of 6 galaxies, found 5 massive galaxies hosting AGNs at $3<z<4$, simultaneously fitting the galaxy and possible AGN contribution to the far-ultraviolet to the far-infrared spectral energy distribution and strong [OIII] emission lines.

Recent observations using the NIRSpec spectrograph on JWST \citep{refId0} have revealed a massive quiescent galaxy, GS-9209, at $z = 4.658$ with an AGN identified by broad H$\alpha$ emission, implying that AGN feedback plays a significant role in quenching star formation \citep{carnall2023massive, Nanayakkara_2023}. The Advanced Deep Extragalactic Survey (JADES) has found 42 type-2 AGN candidates in low mass galaxies up to $z=10$, analyzing UV emission lines with NIRSpec. \citep{Scholtz2023}. Nevertheless, due to the limited field of view of JWST, there have been few spectroscopic observations of massive galaxies at these redshifts to date.\\

S18 presented a sample of massive quiescent galaxies observed with the MOSFIRE
spectrograph on Keck, ten of which were spectroscopically confirmed. Two of these 
galaxies had broad emission lines in their spectra, for which they proposed an AGN origin. In this
paper we consider a larger sample of 22  massive galaxies at $3<z<4$ with
 $\log{(M_\star / M_{\odot}})>10$ and similar
quality spectroscopy. This extends the analysis to include the massive star-forming population at these redshifts.
We will show that broad emission lines are surprisingly common, and we 
explore how these can arise physically.

 This paper is structured as follows. Section \ref{sec:SAMPLEANDDATA} describes the ZFOURGE and multi-wavelength data sets, the sample construction, and emission line fitting.  In Section \ref{sec:ANALYSIS}, we calculate structural properties and consider  a scenario where the broad lines arise from 
 galaxy-wide kinematics. In section \ref{sec:AGN} we present our findings regarding the most plausible explanation for the broad emission lines arising 
 from AGN and discuss the implications in Section~\ref{sec:DISCUSSION}. Finally, in Section~\ref{sec:CONCLUSIONS}, we summarise  the implications of the results.\\
 
 Throughout this paper, we use AB magnitudes and adopt a flat $\Lambda$CDM cosmology with $\Omega_{\Lambda} =0.7$, $\Omega_{M} =0.3$, and $H_{0} =70$ km$s^{-1}$Mpc$^{-1}$.

\section{Sample and data}
\label{sec:SAMPLEANDDATA}
\subsection{Spectral Sample}
\label{sec:SpectralSample} 

Our sample comprises massive galaxies originally identified using photometric redshifts from the medium band near-infrared FourStar Galaxy Evolution Survey  \citep[ZFOURGE;][]{Straatman_2016}. This survey utilizes the near-infrared FOURSTAR instrument on the Magellan telescope across the Chandra Deep Field South \citep[CDFS;][]{Giacconi2002ApJS..139..369G}, COSMOS \citep{Scoville_2007}, and UDS \citep{Grogin2011, Koekemoer2011} legacy fields. FOURSTAR has medium-bandwidth filters from 1--\SI{1.8}{\micro\metre} to calculate robust photometric redshifts, the imaging reaches $5\sigma$ point-source limiting depths of $\sim 25$ AB mag in Ks \citep{Straatman_2016, Spitler_2014}. ZFOURGE is supplemented with existing data from CANDELS Hubble Space Telescope (HST) WFC3/F160W imaging and Spitzer/Infrared Array Camera (IRAC), as well as other ground-based imaging, generating multiwavelength catalogs spanning 0.3–\SI{8}{\micro\metre}.  

We selected galaxies from ZFOURGE with $\log(M_{\star}/M_{\odot})>10$, and
photometric redshift $3<z<4$, and $K<24.5$. We only used the galaxies with a quality flag \texttt{use=1}, obtaining a sample of 130 massive galaxies within the 11'x11' ( total of 450 sq. arcmin) coverage of the  CDFS, COSMOS and UDS fields. 
We search the KOA and ESO archives gathering spectra for 54 of the 130 massive galaxies at $3<z<4$ from MOSFIRE and KMOS spectrographs.

The spectral data was taken from 4 different programs (PI:Glazebrook, Oesch, Illingworth, Kewley), including all the spectra from S18, where the observations, integration time, and data reduction are described. 
The observations were made with the MOSFIRE \citep{IanS} spectrograph installed on the Keck I telescope at the summit of Mauna Kea in Hawaii, and the KMOS spectrograph installed at the VLT at Cerro Paranal in the Atacama Desert in Chile \citep{Davies2013}. MOSFIRE is a multi-object infrared spectrograph with a field of view of 6' $\times$ 3' that can be used to observe up to 46 slits per mask, with a resolving power of R $\sim$ 3500 in a single band from Y to K bands. KMOS is also a K-band Multi-Object Spectrograph capable of performing spectroscopy in the near-infrared band of up to 24 targets simultaneously, with a spectral resolution of R$\sim 3000-4000$. We note our sample includes all galaxy spectra from S18.

All galaxies were observed in the H and K-band on several masks across multiple nights, all observed with standard ABBA exposures, nodding along the slit with a $0.7$ arcsec slit for mask configurations. Each observing program has different integration times in the K band, described in detail in S18. 
MOSFIRE and KMOS have a wavelength coverage of 1.93-\SI{2.45}{\micro\metre} and 2.038–\SI{2.290}{\micro\metre} in the K band. Therefore,  galaxies at z=3.8 reach the detection limit for [OIII]5007 emission lines. The noise also increases at longer wavelengths, and there is a loss due to atmospheric transmission at \SI{2.35}{\micro\metre}. The minimal  detection limit for the [OIII]5007 emission line is about 4.7$\times 10^{-18}$ erg/s/c$m^{2}$.

Out of the 54 galaxies with spectra, we have secured spectroscopic redshifts for 33 galaxies. In Section~\ref{sec:Stackspectra}, we fit a sample of 22 objects with strong [OIII] and H$\beta$ emission lines, which we then analyze further.
Of the remaining 11, 3 lie outside $3<z<4$, and we discard these. A further 5 have redshifts from S18 with weak or no emission lines, and another one has redshifts from weak emission lines or absorption lines, which we determined using {\tt slinefit} \citep{cschreibslinefit}. One further galaxy does not display the  [OIII]5007 emission line necessary for our analysis, showing only H$\beta$ emission line. Lastly, the slit of one of the galaxies is contaminated with a second object and it has been rejected. 21 galaxies have no spectroscopic redshift that can be determined from the spectra.

\subsection{Spectra and Emission line fluxes}
\label{sec:Stackspectra} 

We measure the flux density of the [OIII]5007 and H$\beta$ emission lines with the non-linear least-squares minimization and curve fitting python library \textsc{LMfit-py} \citep{newvillematthew2014}. \textsc{LMfit} allows us to fit all three lines ([OIII]5007, [OIII]4959 and H$\beta$) together, constraining the center and velocity dispersions consistently with redshift, utilizing a Gaussian model centered on the emission line's wavelength. A visual inspection does not find any cases where H$\beta$ emission lines have a different width than [OIII] emission lines. We used the \textsc{Specutils} python package to calculate the standard equivalent width (EW) measurement. We employ a continuum-normalized spectrum to obtain the absorption line width of the adjacent continuum that has the same area as that of the absorption line. 

We fit the emission lines with \textsc{LMFit-py} and remove galaxies from the sample with [OIII]5007 emission line signal-to-noise (S/N) <3. We remove from the sample galaxies with a standard deviation larger than $30\%$ in the full width at half maximum (FWHM) measurements, reducing the sample to 26 galaxies. We focus only on galaxies with detected [OIII]5007 and H$\beta$ emission lines. Therefore we remove four other galaxies (COSMOS-4417, COSMOS-15056, COSMOS-17779, and COSMOS-12000) with H$\beta$4862 absorption lines or which lack detectable H$\beta$ emission lines, further reducing the sample to 22 galaxies. 

Four galaxies in the MOSFIRE archives were covered by more than one spectrum (COSMOS-4214, COSMOS-12173, COSMOS-20169, and UDS-6023) with different slit positions and orientations, which can lead to different velocity profile measurements. Therefore we modeled the spectra jointly, so the H$\beta$ and [OIII] doublet could be identified. 

We show an example 2D and 1D spectrum with the best fit for all the object in each field --- COSMOS, UDS, and CDFS --- in Apendix Figures \ref{sec:Apendix}1-3, with the best fit for the [OIII]4959, [OIII]5007  and H$\beta$4862 lines obtained with \textsc{LMfit}. 


\begin{table}
	\centering
	\caption{Selection properties of our galaxy sample.}
	\label{tab:colors}
	\begin{tabular}{lllll} 
\hline		
CATAID	     &$z_{\textrm{spec}}$ &$K_{\textrm{mag}}$& U-V                & V-J               \\
\hline	
\hline	
COSMOS-2561	 &	3.092 & 22.9    & $1.53^{+0.04}_{-0.03}$ & $1.12^{+0.04}_{-0.04}$\\
COSMOS-4214	 &	3.457 & 23.4    & $1.25^{+0.08}_{-0.07}$ & $1.03^{+0.07}_{-0.07}$\\
COSMOS-5874	 &	3.454 & 22.8    & $1.06^{+0.04}_{-0.03}$ & $1.07^{+0.12}_{-0.02}$\\
COSMOS-12173 &	3.180 & 21.8    & $1.19^{+0.02}_{-0.01}$ & $0.97^{+0.02}_{-0.01}$\\
COSMOS-20133 &	3.481 & 23.8    & $1.23^{+0.04}_{-0.06}$ & $0.72^{+0.04}_{-0.14}$\\
COSMOS-20169 &	3.330 & 24.1    & $1.85^{+0.14}_{-0.07}$ & $1.60^{+0.12}_{-0.07}$\\
CDFS-6691    &  3.473 & 23.0    & $1.56^{+0.02}_{-0.08}$ & $0.90^{+0.04}_{-0.04}$\\
CDFS-7535	 &	3.587 & 21.1    & $0.31^{+0.01}_{-0.03}$ & $0.55^{+0.01}_{-0.03}$\\
CDFS-7752 	 &	3.473 & 22.0    & $0.50^{+0.01}_{-0.05}$ & $0.82^{+0.01}_{-0.08}$\\
CDFS-8428	 &	3.661 & 22.3    & $1.15^{+0.01}_{-0.05}$ & $1.21^{+0.01}_{-0.08}$\\
CDFS-9332	 &	3.704 & 22.1    & $0.97^{+0.02}_{-0.02}$ & $0.94^{+0.07}_{-0.02}$\\
CDFS-9833	 &	3.475 & 22.0    & $0.40^{+0.01}_{-0.07}$ & $0.69^{+0.01}_{-0.13}$\\
CDFS-13954   &  3.062 & 21.6    & $1.29^{+0.02}_{-0.07}$ & $1.12^{+0.05}_{-0.03}$\\
UDS-3417	 &	3.229 & 21.7    & $1.20^{+0.02}_{-0.01}$ & $1.11^{+0.02}_{-0.01}$\\
UDS-6023	 &	3.104 & 23.7    & $1.80^{+0.07}_{-0.05}$ & $1.52^{+0.07}_{-0.05}$\\
UDS-6226	 &	3.196 & 23.0    & $1.11^{+0.01}_{-0.09}$ & $0.84^{+0.12}_{-0.04}$\\
UDS-6639     &  3.071 & 23.3    & $1.57^{+0.05}_{-0.09}$ & $2.24^{+0.6}_{-0.2}$ \\
UDS-8092     &  3.226 & 22.2    & $1.35^{+0.04}_{-0.05}$ & $0.61^{+0.1}_{-0.1}$  \\
UDS-8844	 &	3.198 & 22.9    & $1.20^{+0.01}_{-0.02}$ & $1.41^{+0.02}_{-0.01}$\\
UDS-10644	 &	3.656 & 23.9    & $1.89^{+0.15}_{-0.04}$ & $1.65^{+0.12}_{-0.06}$\\
UDS-14075	 &	3.464 & 23.9    & $0.70^{+0.05}_{-0.12}$ & $0.53^{+0.04}_{-0.14}$\\
UDS-8197     &  3.543 & 23.2    & $1.30^{+0.15}_{-0.12}$ & $0.50^{+0.18}_{-0.02}$ \\
\hline
\end{tabular}
\end{table}

\subsection{Spectral classification of the sample}
\label{sec:Subtracting}

The ($U-V$, $V-J$) rest frame color-color diagram (hereafter UVJ) has become a standard technique
to characterize high-redshift galaxies \citep{Franx_2003, vanDokkum_2003, Williams_2009} and in particular to classify galaxies into quiescent, low obscuration star-forming and high-obscuration star-forming categories based
on the overall shape of the spectral energy distribution (SED). Use of this technique requires extensive multiwavelength photometry and accurate redshifts, which we can find in ZFOURGE.

For our sample, we find galaxies with strong emission lines (see Table \ref{tab:FINALTABLE}); the contribution of these lines to the photometry in the K band, can result in incorrect classifications with UVJ which assumes the broad band photometry traces the stellar continuum. To correct for this we follow the procedure from S18 to correct the photometry for emission line contamination. Equation \ref{eq:SBB} below, assuming a constant continuum flux density within the filter and a constant filter response over the spectral extent of the lines:

\begin{equation}
    \frac{S_{BB}}{S_{BB}^{\textrm{cor}}} = 1 + \sum_{l} EW^{l}_{\textrm{obs}} \frac{R(\lambda_{l})}{\int d(\lambda)R(\lambda)}
	\label{eq:SBB}
\end{equation}

Where $S_{BB}$ is the original flux density, $R(\lambda)$ is the response curve of the corresponding filter, $EW^{l}_{\textrm{obs}}$ the observer-frame equivalent width, and $\lambda_{l}$ the central wavelength of the line $l$.

\begin{figure*}
	\includegraphics[width=\textwidth, height=9cm]{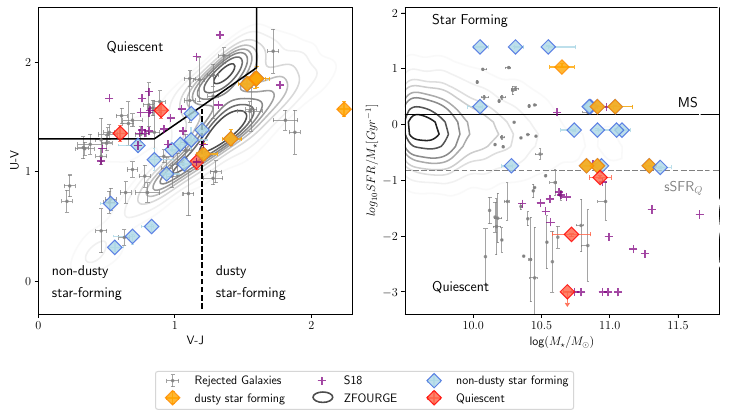}
    \caption{Left: UVJ colour-colour plot. Purple crosses are massive quiescent galaxies at $3<z<4$, from S18. Grey dots are the first selection of galaxies with spectra from MOSFIRE and KMOS archives within ZFOURGE. The diamonds are the emission line galaxies with flux correction (see section \ref{sec:Subtracting}). Red, orange, and blue correspond to quiescent, dusty, and non-dusty star-forming galaxies correspondingly. We have two galaxies in common with S18, identified as AGN candidates. Right: Logarithm of the specific star formation rate (sSFR) versus stellar mass. The black line indicates the main sequence (MS) for galaxies in the ZFOURGE calculated by S18. The entire ZFOURGE galaxy sample is shown as dark contours. The dashed grey line indicates the quenching limit (sSFR$_{Q}$) defined by S18.}
    \label{fig:UVJ}
\end{figure*}

In practice, the emission lines only impact the K band, and after correcting the photometry, all our galaxies dropped in flux, with a median offset of 32 $\%$ and CDFS-9833 being the most affected, with 74$\%$ of its flux removed. In Figure \ref{fig:UVJ} we show the UVJ diagram of the final sample after the emission lines from the broadband photometry were subtracted. It
can be seen that our galaxies mostly fall in the star-forming region of the diagram (both obscured and unobscured); we can also observe the two galaxies that overlap with S18.

In Figure~\ref{fig:UVJ}, we show the distribution of the specific star formation rate (sSFR) of our final galaxy sample and compare it to the massive quiescent galaxies from S18. Star-forming galaxies show an empirical relation between specific star formation rates and stellar mass, widely known as star-forming main sequence (MS) \citep{Elbaz2007A&A...468...33E,Noeske2007}.
S18 uses a definition for `quiescence' of sSFR=sSFR$_{Q}=0.15 $Gyr$^{-1}$, valid for galaxies $z>3$, and correspond to a `relative' quiescence, as galaxies below this threshold at $z>3$ are forming stars with rates an order of magnitude lower than the bulk of the star-forming population. We can observe that most of the galaxies in our sample reside close to the MS in comparison with the MS from ZFOURGE and above the sSFR$_{Q}$ except for 3 galaxies. One of them is UDS-8197 from S18, that analysis proposed as an AGN candidate from the emission lines in its spectra. One of the other two galaxies UDS-8092 was detected as an AGN in X-rays by \cite{Cowley2016} and the final one CDFS-6691, even though it appears quiescent in the UVJ plot and lower than sSFR$_{Q}$, shows [OIII]5007 emission lines, with low H$\beta$ emission. We will keep displaying these galaxies in red to be able to trace them during the development of the paper.

\subsection{Structural Properties}
\label{sec:Structural}

We measure the half-light radius along the semi-major axis $r_{e}$ and Sersic index $n$ of our galaxy sample from H-band WFC3/F160W HST images, utilizing the profile modeling software \textsc{GALFIT} \citep{Peng2002}. We use single components Sersic profiles, providing an initial $r_{e}$, Sersic index, and axis ratio $q=b/a$. Point Spread Function (PSFs) were taken from \citep{Skelton2014, Grogin2011, Koekemoer2011} for COSMOS, GOODS, and UDS fields. 

We compare our measurements obtained with \textsc{GALFIT} to the galaxies with a good fit flag from CANDELS/3D-HST catalogs \citep{vanderWel_2014}, finding a good agreement (2-3$\%$). We find four galaxies with bad fit flags in our sample (COSMOS-20133, UDS-6023, CDFS-7535 and CDFS-9332), where inputs for axis ratio (b/a) and position angle (PA) from \citep{vanderWel_2014} were inaccurate, creating poor models and therefore unphysical $r_{e}$ and Sersic index for those galaxies. We corrected those galaxies that resulted in a bad fit and summarized the results in Table \ref{tab:FINALTABLE}. We also noticed no sign of multiple sources or merging systems in the GALFIT analysis. 

\subsection{Stellar Masses}

The CANDELS/3D-HST catalogs also provide stellar masses for our galaxy sample. These masses were calculated with photometric redshifts, and  do not take for account the presence of AGN in the galaxies, which could originate the emission lines that we observe in the spectra. Therefore, we re-calculated them with the new spectroscopic redshifts (see Table \ref{tab:FINALTABLE} and ZFOURGE photometry).\\

We use the code \textsc{CIGALE} \citep{Boquien2019} to re-compute the stellar masses, utilizing the \citep{Bruzual2003} stellar population model. We assume a \citep{Chabrier_2003} initial mass function (IMF) with a solar metallicity ($Z = 0.02$). For the dust attenuation, we adopt {\tt dustatt\_modified\_starburst} in CIGALE \citep{Calzetti_2000} with $A_V$ up to 6 mag, and for the dust emission models we use \citep{Casey2012}. For star formation history (SFH), we use a delayed SFH with optional exponential burst (except for galaxy UDS-3417, which uses a delayed SFH with an optional constant burst/quench in order to find a suitable fit with a $\tilde{\chi}^2<2$), we allow the e-folding time and stellar age varying from $0.3$–-1 Gyr to $0.5$-–2 Gyr, respectively. For the AGN component, we adopt the {\tt skirtor2016} module based on a clumpy torus model SKITOR AGN from \citep{Stalevski2012, Stalevski2016}. The relative strength between the AGN and galaxy components are set by the {\tt fracAGN} parameter, we allow {\tt fracAGN} to vary from $0.1$ to $0.9$, and the 9.7$\mu$m optical depth to be 3, 7 and 11. \\ The stellar masses calculated with CIGALE show a median offset of 0.242 and a DEX scatter of 0.14 compared to the CANDELS/3D-HST catalogue. (See Table \ref{tab:FINALTABLE}).

\section{Possible origins of broad lines}
\label{sec:ANALYSIS}
After correction for emission line contributions, most of our galaxies lie in the star-forming regions, both obscured and unobscured, of the UVJ diagram. Broad emission lines are common with 13/22 having velocity dispersions $>200$ km$s^{-1}$ (with several $>400$ km$s^{-1}$ and rest-frame equivalent widths up to 2700 \AA; see Table \ref{tab:FINALTABLE}). Such a high frequency of strong, broad emission lines are not typically
seen in lower-redshift samples of massive galaxies \citep{PagottorefId0, Freeman_2019}. 

Broad lines in galaxies can arise due to two principle mechanisms. The first is the kinematics of ionized gas distributed throughout a galaxy and moving in its gravitational potential. For example, the integrated line width of emission lines was used to study the Tully-Fisher scaling relations in high-redshift galaxies \citep{glazebrook_2013, Christensen10.1093/mnras/stx1390} before the advent of integral field spectroscopy. In this dynamical scenario, we would expect to be able to estimate an implied dynamical mass from the galaxy size and velocity width, and this should show a reasonable correlation with stellar mass \citep[e.g.][]{Straatman_2022, Kriek2009}. There is a basic underlying assumption here of virial equilibrium conditions; either pressure or rotationally supported.
Given the sizes of our galaxies and that the spectroscopy is with ground natural seeing it is not possible to determine the spatial
extent of the emission line flux; so we use the HST continuum sizes measured in subsection \ref{sec:Structural} instead. We note that our line widths are much larger than
typically seen in low-redshift kinematic surveys \citep{Franx_2008}. Nevertheless, some of our galaxies
are very compact, and that can give rise to comparable large dispersions in the emission lines \citep{vanDokkum08, Nelson2014}.

The second common mechanism we consider is that the emission comes from a small region at the galaxy center due to the accretion of matter by a supermassive black hole, i.e. an AGN.
Our line widths are comparable with those seen in Type 2 AGN\footnote{While we have been
calling our emission `broad' by comparison with normal star-forming galaxies they are in the range
of the `narrow line regions' of AGN and not the broad line region which has FWHM $>2000$ km/s.}. We can test this
scenario by analyzing the ionization mechanism using line ratios. 
This is commonly done using the 
`Baldwin, Phillips $\&$ Terlevich' (BPT) diagram \citep{BPT}, which can classify galaxies into star-forming or AGN, from their emission-line intensity ratios. This is extensively used
to diagnose ionization mechanisms in galaxies; in particular, AGN are strongly
enhanced in [OIII]$/$H$\beta$ and mildly enhanced in  [NII]/H$\alpha$ which serves to distinguish
strong AGN from star-forming galaxies \citep{Kauffmann2003, Heckman2004ApJ...613..109H, Yan2006}. In our case the redshifted [NII]/H$\alpha$ lies outside the red-end of our ground-based spectra so instead we utilize the mass-excitation (MEx) diagram \citep{Juneau2011}. The MEx diagram leverages the relationship between stellar mass and metallicity in place of the [NII]/H$\alpha$ ratio. By measuring the H$\beta$ and [OIII] ratios, we can identify AGNs through the MEx-diagram. We note that we 
 may still  find a weak correlation between a dynamical mass (calculated under the galaxy-wide
 assumption) and stellar mass, as the velocity dispersion from the [OIII] line will trace the size of the narrow line region (NLR), which extends up to a few parsecs, as its radius is proportional to the square root of the [OIII] luminosity \citep{Bennert_2002, Schmitt2003} and hence be correlated with the mass of the SMBH which also correlates with stellar mass \citep{Kormendy_2013}.

\subsection{Exploring a dynamical origin}

In a dynamical scenario, the emission lines in our galaxy spectra would trace the kinematics of ionized gas distributed throughout the galaxy. In this case, we can expect a strong correlation between the estimated dynamical masses and the stellar masses \citep{Schreiber2018, esdaile2021}. Thus, star formation would be enough to explain the emission lines in our galaxies.  \\

We use the $\sigma_{e}$ and the F160W band $r_{e}$ measurements to calculate `dynamical masses' with the following equation (\ref{eq:4}):

\begin{align}
 &M_{\textrm{dyn}}=\frac{\beta(n)\sigma_{e}^{2}r_{e}}{G}
 \label{eq:4}
\end{align}

Where $\beta(n)$ is the virial coefficient used to approximately account for structural and orbital 
non-homology \citep{Cappellari2006}:
\\
\begin{equation}
    \beta(n) = 8.87 - 0.831n + 0.024n^{2};
	\label{eq:5}
\end{equation}
\\
$r_{e}$ is the half-light radius, $\sigma_{e}$ is the line of sight velocity dispersion (LOSVD) within the $r_{e}$ and G is the gravitational constant. This equation essentially
assumes virial equilibrium and has been commonly used in high-redshift kinematic papers \citep{Cappellari1310.1093/mnras/stt562, Schreiber2018,esdaile2021}.

The derived dynamical masses are in the range $9 < \log_{10}({M}/{M_{\odot}}) < 11.6$. 
We note that some of these values are rather high, and in the light of our conclusions below
favoring the AGN scenario they are not likely to represent true dynamical masses.\\

After obtaining spectral classification, stellar mass, effective radius, velocity dispersion, and dynamical mass for our galaxies, we analyze our sample and compare it with those of massive galaxies at various redshifts.

\begin{figure*}
	\includegraphics[width=\textwidth]{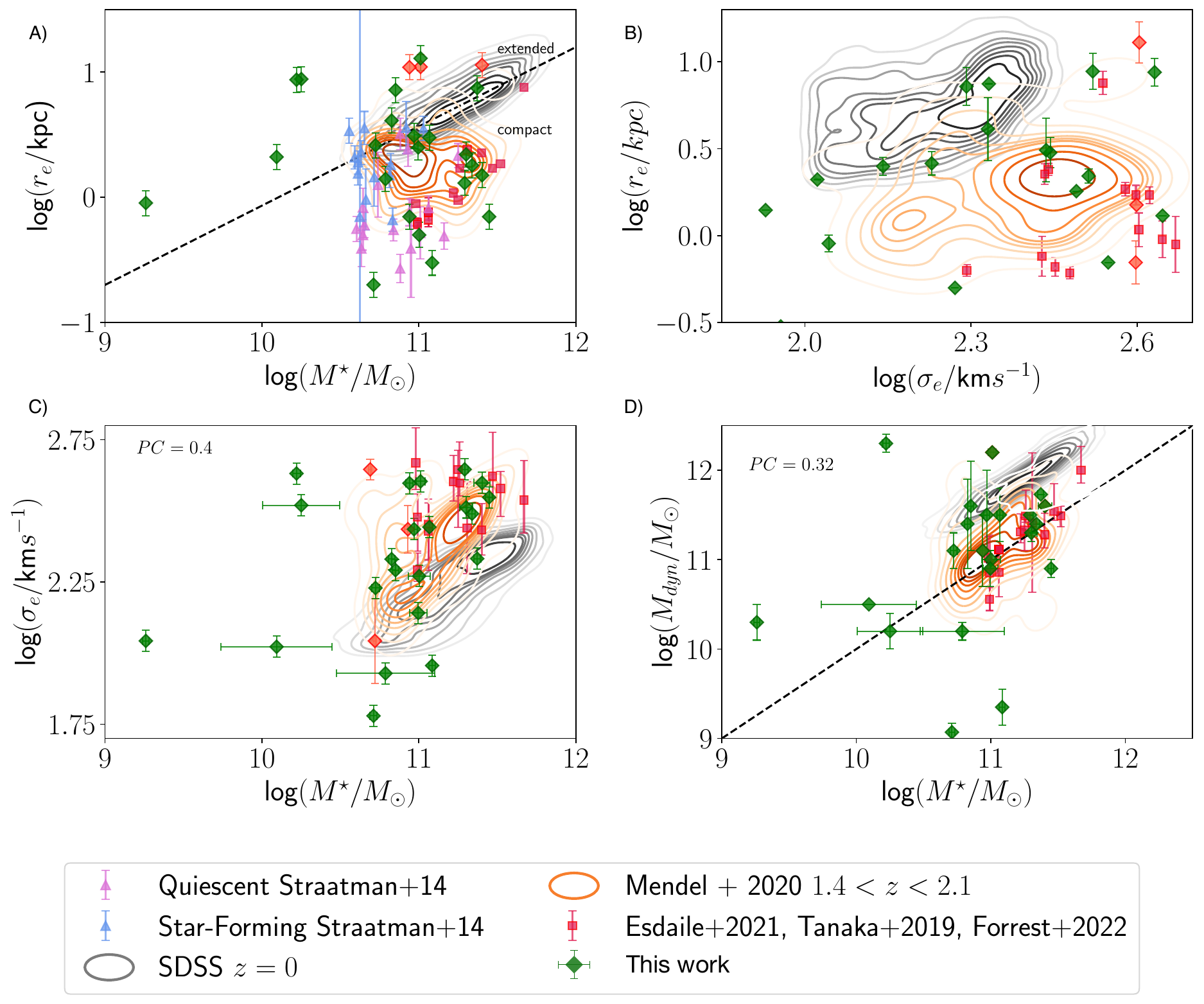}
    \caption{Structural and dynamical comparison of massive galaxies. We compare $r_{e}$, $M_{\star}$ and $\sigma_{e}$ for a range of redshifts. Dark contours are massive galaxies from the SDSS with $z\sim0$. Orange countours show galaxies from \protect\cite{Mendel_2020} at $1.4<z<2.1$. Red squares show QGs from \protect\cite{Tanaka_2019}, \protect\cite{esdaile2021} and \protect\cite{Forrest2022}, at $3<z<4$. The blue and pink triangles are star-forming and quiescent galaxies from \citep{Straatman2014} respectively. Green diamonds are star-forming galaxies from our sample at $3<z<4$.}
    \label{fig:FigureIII}
\end{figure*}

We examine the size--velocity--stellar mass scaling relations; and then compare the implied dynamical mass derived from size and velocity to stellar mass. Our principle comparison is with other samples of massive galaxies at redshifts $3<z<4$ with kinematic analysis that have a good correlation between stellar and dynamical mass. In particular in Figure \ref{fig:FigureIII} we compare our galaxy sample with galaxies at $z\sim 0$ from SDSS \citep{Maraston2005, Thomas2005}, galaxies from \cite{Mendel_2020} at $1.4<z<2.1$, and quiescent massive galaxies from \cite{esdaile2021}, \cite{Forrest2022} and \cite{Tanaka_2019} at $3<z<4$. 

Considering size-mass relation, our analysis indicates that the galaxy sample has an average effective radius of r$_{e}$=2.9$\pm$0.15 kpc, as shown in Fig \ref{fig:FigureIII}-a and \ref{fig:FigureIII}-b. Interestingly, this finding aligns with \cite{Straatman2014}, who observed that star-forming galaxies are 3.2$\pm$1.3 times larger than quiescent galaxies (r$_{e}$=0.63$\pm$0.18 kpc) at z$\sim$4. 
Notably, we also detected a wide range in size, and poor correlations of size vs mass and size vs velocity, with some galaxies exhibiting similar effective radii as those at lower redshifts and others as large as those at z$\sim$0. This suggests that the diversity of the galaxy population was established early in the universe. Overall our galaxy sample extends to lower stellar masses than the QGs at the same redshift range, we interpret this as a selection effect as absorption line kinematic studies can only be done for the brightest and most massive objects. \\

Comparing the velocity dispersion and the stellar mass of the galaxies, we find a weak correlation with a Pearson correlation factor of $0.4 \pm 0.01$ after 1000 Monte Carlo sampling. We note that the comparison samples measured velocity dispersions using absorption lines, whereas we used emission lines. However, these should be suitable for testing a galaxy-wide kinematic origin of such lines. In low redshift star-forming galaxies, emission and absorption largely trace the same kinematics \citep{Cortese_2016}. 

In the dynamical origin hyopthesis, the velocity dispersion of the emission lines detected will trace the kinematics of ionized gas produced by star formation distributed throughout a galaxy due to its gravitational potential. In this scenario, we should expect an agreement between the stellar and dynamical mass of the galaxy.
Comparing the dynamical mass with the stellar mass (see Fig \ref{fig:FigureIII}-d), we found a weak Pearson correlation factor of $0.29 \pm 0.01$ after 1000 Monte Carlo sampling.

This weak correlation could arise from the similar weak correlation of velocity dispersion and stellar mass which could also arise in the AGN scenario (see Section \ref{sec:Emissionlined}). To
understand this better, we need to see if the combination of size and velocity dispersion which goes into the dynamical mass calculation, provides more information than from velocity alone. To test this, we ran 1000 Monte Carlo simulations where we replaced the size of each galaxy with a size value picked randomly from the full sample and recomputed the dynamical mass. Doing this, we found a correlation coefficient of $0.32 \pm 0.02$. This is not significantly different from the values using the true size, so we conclude that the weak correlation we see is simply due to the velocity--stellar mass correlation, which can also arise under the AGN scenario.

Putting this all together we conclude that the velocity dispersion of the [OIII]$_{5007}$ emission line does not appear to be tracing the gravitational potential of the galaxy, thus is unlikely to be due (solely) to star formation.

\section{The presence of AGN in our sample}
\label{sec:AGN}

AGNs emit radiation across the entire electromagnetic spectrum, so in order to further test the AGN hypothesis, we first look for counterpart detections of our galaxies at other wavelengths. We cross-match our galaxy sample with the \cite{Cowley2016} AGN catalog (using a tolerance of $\pm  1$ arcsec), which considers infrared, X-ray, and radio AGN detections in the ZFOURGE sample. Only one of the galaxies in our sample is detected as an infrared AGN in this catalog (COSMOS-12173), with a counterpart detected in X-rays. We find one galaxy detected in both X-ray and radio (CDFS-13954) and three detected only in X-ray (CDFS-9332, CDFS-8428 and UDS-8092). Two QGs were earlier identified as AGN candidates by S18 (COSMOS-20133, UDS-8197), due to their significant [OIII] emission. In summary, we only have counterpart observations that indicate the presence of AGNs for 5 of our 22 galaxies. 

\subsection{Emission line diagnostic diagrams}
\label{sec:Emissionlined}
By analyzing certain emission line intensity ratios, it is possible to determine whether the ionized regions belong to a galaxy that is actively forming stars or if the lines are being powered by a more intense ionizing spectrum linked to an AGN \citep{Kewley_2001, Kauffmann2003, Brinchmann2004, Kewley2008}.
The classic BPT emission line diagnostic diagram, compares intensity ratios of line luminosities [OIII]/H$\beta$ and [NII]/H$\alpha$. 
Near infrared spectra in the K band cover [NII]6583 out to $z=2.5$ and as our galaxy sample is at $z>3$, [NII]6583 is redshifted beyond our observational window. Therefore, we can not use the BPT diagram to distinguish between AGN and purely star-forming galaxies. However, we can calculate the ratios of [OIII]5007/H$\beta$ to use the empirical equivalent for high redshifts, the mass excitation plot \citep[MEx;][]{Juneau2011}. 

The MEx plot replaces the ratio of the redder emission lines ([NII]6583/H$\alpha$) with the stellar mass to fill gaps where these lines fall in especially noisy regions or outside of the atmospheric cut-off. This diagram utilizes a probabilistic approach to split galaxies into sub-categories and finds intrinsically weak and heavily absorbed AGNs, taking advantage of the known stellar mass–metallicity (MZ) relation. One might expect the MEx diagram to be more sensitive to changes in the stellar-mass–metallicity (MZ) relation compared to the traditional BPT.

 \cite{Juneau2011} presented the MEx as an AGN diagnostic tool applicable to galaxies out to $z=1$. However they explored possible evolutionary effects, and they found some evidence that the intermediate-redshift galaxies may be offset in the MEx diagram relative to the lower-redshift sample (\citealt{Juneau_2014}, J14 hereafter;  \citealt{Coil_2015, Strom_2017}).
  These differences can be attributed to variations in the underlying stellar populations, as well as differences in the interstellar medium conditions (ISM). J14 shifted the original criterion by $\Delta$log(M$_{\star}$/M$_{\sun}$)=$0.25$ expanding the capabilities of the MEx diagram up to $z=2$. 
   
\begin{figure*}
    
\centering
	\includegraphics[width=\textwidth]{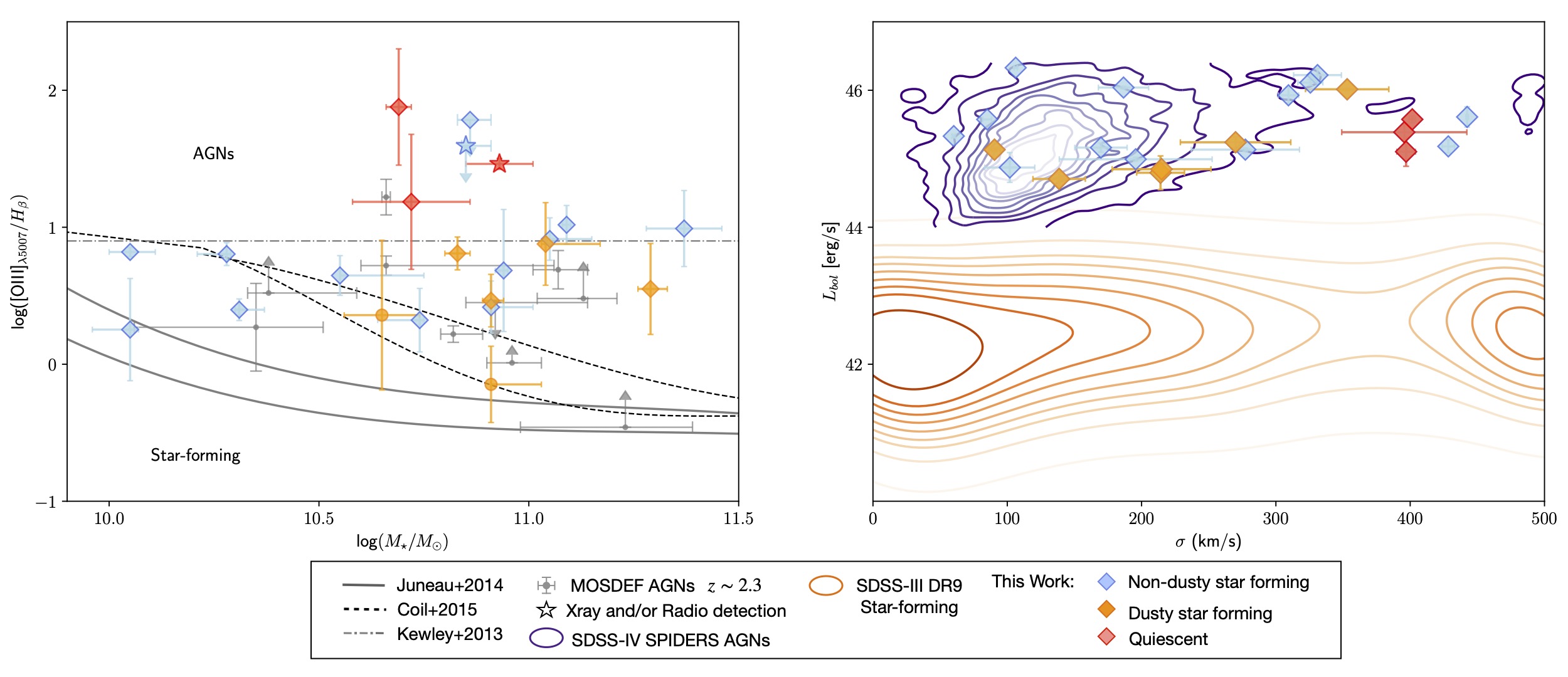}
    \caption{Light blue diamonds are non-dusty star-forming galaxies, orange diamonds are dusty star-forming galaxies and red diamonds are quiescent galaxies from our sample. On the left side, we show the Mass excitation (MEx) plot. The grey lines are the limits of the J14 MEx plot, the black dashed line is the C15 criteria for the MEx diagram. The dot-dashed line corresponds to the theoretical BPT ratio for [OIII]/H$\beta$ from K13. The stars represent the galaxies detected either in X-rays or radio. On the right side, we plot the velocity dispersion against the $L_{bol} \sim 454 \times L_{OIII}$ of the galaxies. The orange contours are star-forming galaxies from the SDSS-III DR9 at $0<z<0.75$ and the purple contours are AGN galaxies from the SDSS IV SPIDERS at $0<z<2.5$. }
    \label{fig:mass_exitation_colors}
\end{figure*}

In Figure \ref{fig:mass_exitation_colors}, we show the log([OIII]/H$\beta$) ratios versus the stellar mass of our galaxy sample. We plot with a black line the J14 limits. Based on this limit, all galaxies in our sample are AGN candidates, with only one galaxy (UDS-8844) likely to be in the MEx intermediate region, indicating both star formation and an AGN.\\

 The boundary between AGN and star-forming galaxies in emission line diagnostics diagrams is fuzzy; \citep{Kewley_2013} (hereafter K13) used a theoretical approach to make predictions about the locations of galaxies and AGNs in the BPT diagram up to $z=3$. They considered four different evolutionary scenarios, which involved changing the ISM conditions (ranging from normal to extreme) and the metallicity (from metal-poor to metal-rich) of the AGN narrow line region at high redshift. An ``extreme" ISM condition could result from a larger ionization parameter, a denser ISM, or a harder ionizing radiation field. In a metal-rich scenario, the gas near the AGN is enriched to a higher metallicity than is typically found in the host galaxy, while in the metal-poor scenario, the gas near the AGN has the same metallicity as the gas in the host galaxy. Regardless of the scenario, all predictions for galaxies at $z=3$ indicate that 
 sources with $\log_{10}( \rm [OIII]/H\beta)>1$ are ionized by AGN, but there is a lack of observational data to test these predictions at this redshift. \\
 
\citep{Coil_2015} (hereafter C15) tested the MEx diagram with a sample of galaxies identified as AGNs either through X-ray or IR emission at $z\sim2.3$ from the MOSFIRE Deep Evolution Field survey (MOSDEF) sample \citep{Kriek2015}. They propose a substantially higher shift of $\Delta$log(M$_{\star}$/M$_{\sun}$)=$0.75$ for the J14 limits. We have included a dashed line on the graph to represent the stricter classification criteria used in C15. According to this approach, three galaxies (CDFS-9833, UDS-14075, CDFS-7752) would be considered star-forming, and the other three galaxies (UDS-8844, COSMOS-4214, UDS-6639) would fall into the MEx mixing sequence. Galaxies along the mixing sequence are called composites or transition objects. They may have a mix of star formation, shock excitation, and/or AGN activity. However, we also plot with grey dots the AGN sample from C15, and two galaxies with a log(M$_{\star}$/M$_{\sun})<10.5$, confirmed as AGNs with X-rays, which are under the C15 limits (Fig. \ref{fig:mass_exitation_colors}). 
We also show with a dot-dashed line the theoretical limit from K13, noticing that the five X-ray detected galaxies in our sample (plotted with a star symbol) are above this limit, located in the AGN region of the MEx diagram. 
The line ratios diagnostic methods can be sensitive to metallicity and can move the position of the AGN to lower metallicities  \cite[lower masses in the MEx diagram;][]{Kewley_2013}. Although low-metallicity AGNs are extremely rare among local galaxies, we are dealing with a sample of galaxies at high redshift, which can be an important factor in explaining the presence of our galaxies in this mixing sequence \citep{harikane2023jwstnirspec}.\\

This  galaxy sample could be a missed population of heavily obscured AGN, or the  X-ray emission is too faint to be detected at this redshift \citep{Straatman2014, Cowley2016, Forrest2020}. 

\subsection{[OIII] luminosities and bolometric correction}

We can calculate the bolometric luminosity of our galaxy sample to compare them with galaxies that are known to host an AGN at lower redshifts and see if they have similar luminosity ranges. To obtain the bolometric luminosity of our galaxy sample, we used the [OIII]5007 luminosity and applied the luminosity-dependent [OIII] bolometric correction factor, $C_{\textrm{OIII}}$ from \cite{Lamastra_2009}, $L_{bol} \sim 454 \times L_{\textrm{OIII}}$, assuming that the observed [OIII] line luminosities are a lower limit to the intrinsic luminosity of [OIII].

From Table \ref{tab:FINALTABLE} we can see that the luminosities range from $5 \times 10^{44}$ erg s$^{-1}$ to $2 \times 10^{46}$ erg s$^{-1}$, which is consistent with them hosting powerful hidden AGNs. In Figure \ref{fig:mass_exitation_colors} we compare the bolometric luminosity ($L_{\textrm{bol}}$) of our galaxy sample with the $L_{\textrm{bol}}$ from the catalog of optical spectral properties for all X-ray selected SPIDERS AGNs at redshifts $0<z<2.5$ \citep{Coffey2019} and star forming galaxies from SDSS-III DR9 spectra between $0<z<0.75$ \citep{Brinchmann2004, Kauffmann2003, Tremonti2004ApJ}. The SPIDERS AGN and the SDSS-III DR9 catalogues, estimated the bolometric luminosities using the bolometric corrections from \citep{Richards2006} and \citep{Shen2011} for $L_{3000 \angstrom}$ and $L_{5100 \angstrom}$.
We utilize the $[OIII]_{5007}$ luminosities from the value-added catalogs from the MPA-JHU \citep{Brinchmann2004, Kauffmann2003, Tremonti2004ApJ} to calculate the bolometric luminosities of the star-forming galaxies of the SDSS-III DR9 catalogs.
From the plot, it is evident that the luminosity distribution of our galaxy sample is greater than that of the star-forming galaxy region. Moreover, our sample lies at the upper end of $L_{\textrm{bol}}$ of the AGN region. As shown in other works, this can provide further evidence that our sample galaxies likely host AGN \citep{Marsan2017ApJ, Harrison2016}.\\

The radiative transfer code SKITOR \citep{Stalevski2016}, implemented in CIGALE, can model the distribution of dust in the torus as a two-phase medium. This allows us to estimate the infrared radiation emitted by the dust to obtain the ratio of the torus to the AGN luminosity, as well as the fraction of AGN IR luminosity to total IR luminosity ($f^{IR}_{AGN}$). We estimated the $f^{IR}_{AGN}$  from the SED fitting of our galaxy sample (see Table \ref{tab:FINALTABLE}) and obtained a $f^{IR}_{AGN} > 0.2$ for the entire sample. A $f^{IR}_{AGN}<0.2$ is considered a non-dominant AGN, while a $f^{IR}_{AGN}> 0.85$ is a highly-dominant AGN \citep{Ciesla2015, Ramos_Padilla_2021}. Overall, the galaxies in our sample show an important  $f^{IR}_{AGN}$, supporting the presence of AGN in our galaxies.

\subsection{AGN Fraction}

The fraction of AGNs ($f_{\textrm{AGN}}$) within a galaxy population is an indicator of AGN activity and supermassive black hole growth at galaxy centers. The energy output of AGNs is a powerful feedback mechanism that can effectively regulate gas inflow and star formation in large galaxies. As such, assessing the level of AGN activity in massive galaxies with respect to redshift is a valuable means of evaluating the impact of AGN feedback at any cosmic epoch. The $f_{\textrm{AGN}}$ for massive galaxies with log(M$_{\star}$/M$_{\sun})$>11 seems to evolve across cosmic time, as different surveys show that $f_{\textrm{AGN}}$ changes with redshift, showing a dramatic transition in the $f_{\textrm{AGN}}$  happening at $z\sim2.5$ when the universe was $\sim$3Gyr old \citep{Kauffmann2003, Kriek2009, Marsan2017ApJ}.

We calculate the fraction of AGNs in our sample assuming all of our galaxies are AGNs, utilazing the J14 limits of the MEx diagram as follows:

\begin{equation}
    f_{\textrm{AGN}}=\frac{\text{Number of AGN}}{\text{Number of galaxies}}
	\label{eq:6}
\end{equation}

\begin{figure}
\centering
\includegraphics[width=\columnwidth]{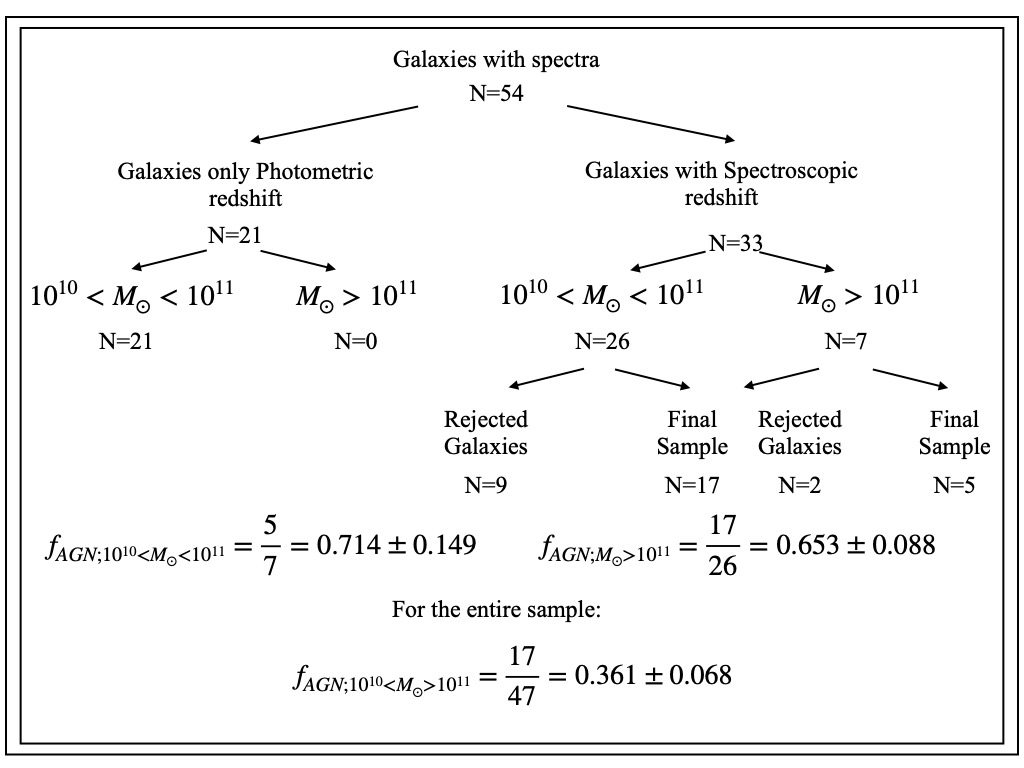}
    \caption[width=\columnwidth]{Summary Schema of the $f_{\textrm{AGN}}$. We use a beta distribution to calculate the error bars.}
    \label{fig:AGNfschemma}
\end{figure}

Where the numerator is the number of AGN detected by broad [OIII] and high [OIII]/H$\beta$ within $3<z<4$, and the denominator, the number of galaxies with spectra with spectroscopic redshift detected within $3<z<4$ (See Figure \ref{fig:AGNfschemma}). We use the beta distribution \citep{Gupta2011} to derive errors on the ratios. We note that our sample may be biased towards the strongest emission lines as these are easier to detect and that we have 21 galaxies with unconfirmed spectroscopic redshifts between $3<z<4$. We summarise the $f_{\textrm{AGN}}$ in Table \ref{tab:fagn}.

\begin{figure}
\centering
\includegraphics[width=\columnwidth]{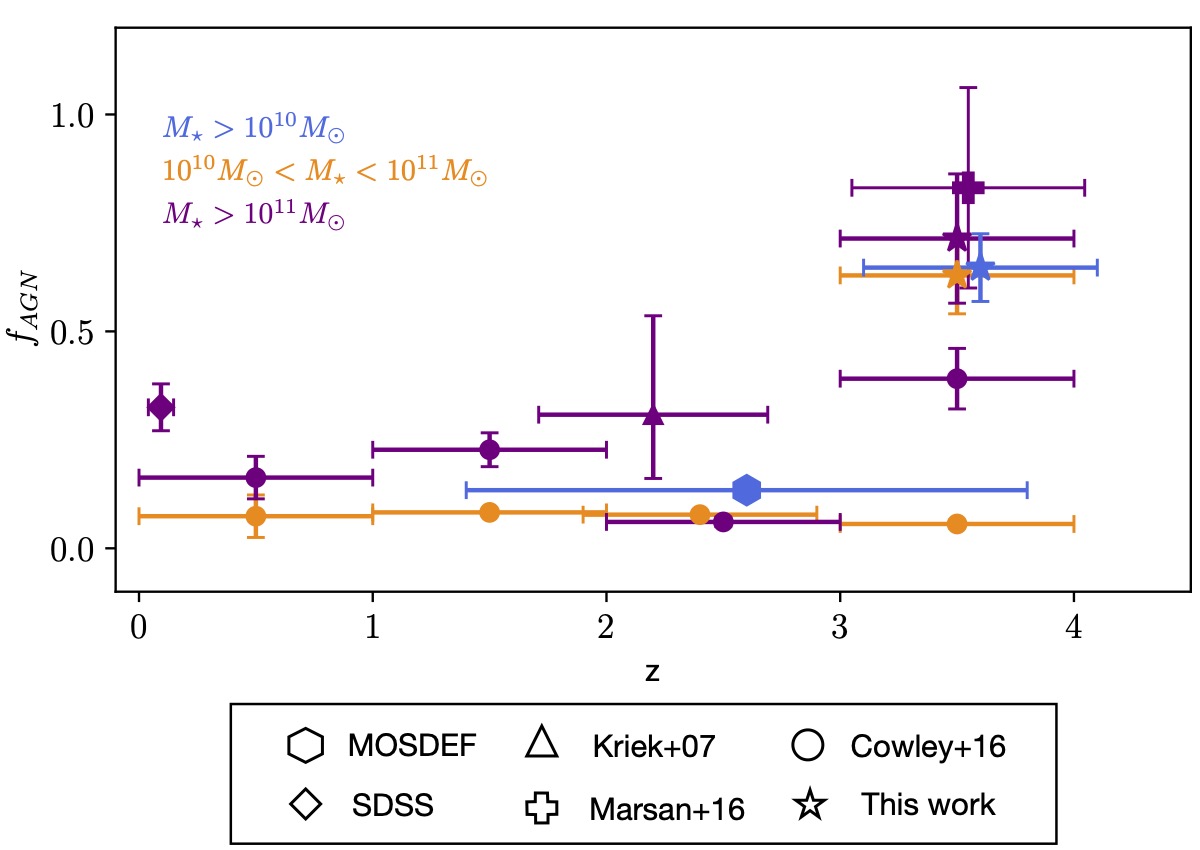}
    \caption[width=\columnwidth]{ AGN fraction evolution between $0<z<4$. Blue, orange and purple colors, differentiate the different stellar mass ranges. The dots show the $f_{AGN}$ for ZFOURGE \cite{Cowley2016}, the diamond show the $f_{AGN}$ for SDSS \cite{Kauffmann2003}, the triangle show the $f_{AGN}$ for \cite{Kriek2009}, the hexagon show the $f_{AGN}$ for MOSDEF, the cross show the $f_{AGN}$ \cite{Marsan2017ApJ} and the stars show the $f_{AGN}$ of this work.}
    \label{fig:fAGN}
\end{figure}

In order to gain a better understanding of the $f_{\textrm{AGN}}$, it is important to compare our results to similar stellar mass ranges and also see the evolution of the $f_{\textrm{AGN}}$ with cosmic time. So, we compare our $f_{\textrm{AGN}}$ in Figure \ref{fig:fAGN} to different samples, in three stellar mass ranges (log(M$_{\star}$/M$_{\sun})$>10, 10<log(M$_{\star}$/M$_{\sun})$<11 and log(M$_{\star}$/M$_{\sun})$>11), and to different $f_{\textrm{AGN}}$ between $0<z<4$. 

Our sample comes originally from ZFOURGE, therefore we also calculate the $f_{\textrm{AGN}}$ for galaxies with stellar mass 10<log(M$_{\star}$/M$_{\sun})$<11 and log(M$_{\star}$/M$_{\sun})$>11 of the ZFOURGE catalog at different redshift ranges to trace the cosmic evolution of the $f_{\textrm{AGN}}$ up to $z=4$. Nevertheless, as the ZFOURGE catalog provides AGN from multi-wavelength detections, this time the numerator is the number of Xray+FIR+Radio AGN detections, and the denominator is the number of objects in the ZFOURGE catalog with flag \texttt{use=1}. We summarize the AGN fraction in table \ref{tab:fagn}.

Our sample is primarily made up of MOSFIRE spectra, which allows us to compare our AGN fraction to that of the MOSDEF survey. \cite{Aird_2017} utilize a multiwavelength approach of optical, IR, and Xray data to find the AGNs in the MOSDEF sample. They identified 55 AGNs out of their 482 galaxy sample, where 24 AGNs were detected by the line rations of the BPT diagram obtained from MOSFIRE spectra. The MOSDEF sample covers galaxies from $1.4 < z < 3.8$, nevertheless, most of the optical AGNs are at $2<z<3$. So, we just compare a rough $f_{\textrm{AGN}}=0.134 \pm 0.009$ for galaxies between $1.4 < z < 3.8$ and log(M$_{\star}$/M$_{\sun})$>10 (in blue colors) in Figure \ref{fig:fAGN}. 

For galaxies with 10<log(M$_{\star}$/M$_{\sun})$<11, we compare (in orange colours) our $f_{\textrm{AGN}}=0.629 \pm 0.0885$ to the $f_{\textrm{AGN}}$ from ZFOURGE at different redshits. 

Finally, for galaxies with log(M$_{\star}$/M$_{\sun})$>11, we compare (in purple colours) our $f_{\textrm{AGN}}=0.714 \pm 0.149$ to \citep{Marsan2017ApJ} with $f_{\textrm{AGN}}=0.83\pm0.231$ in the same redshift range $3<z<4$. \citep{Marsan2017ApJ} analyze a sample of 5 galaxies, where 4 of them were identified as AGNs from strong [O III] emission lines in the IR SED properties and X-ray and radio detections. We also compare our AGN fraction to the ZFOURGE catalog  $f_{\textrm{AGN}}=0.391 \pm 0.0698$. 
We also add the $f_{\textrm{AGN}}$ from SDSS \citep{Kauffmann2003} at $z=0$ and the $f_{\textrm{AGN}}$ from \citep{Kriek2009} at $1.7<z<2.7$, to trace the $f_{\textrm{AGN}}$ through redshift.

\begin{table}
	\centering
	\caption{AGN galaxy fraction of ZFOURGE.}
	\label{tab:fagn}
	\begin{tabular}{|ll|c|c|} 
\cline{3-4}
 \multicolumn{2}{c|}{} & \multicolumn{2}{|c|}{$f_{AGN}$} \\
\hline
z                       &   Sample          &  $10^{10}$<$M_{\odot}$<$10^{11}$  & $M_{\odot}$>$10^{11}$  \\
\hline
\hline
\multirow{2}{5em}{0<z<1}& ZFOURGE &   $0.074\pm 0.049$   & $0.163 \pm 0.049$ \\
                        & SDSS    &                      & $0.325\pm 0.054$  \\
\hline
1<z<2                   & ZFOURGE &   $0.082\pm 0.006$   &  $0.222 \pm 0.039$ \\
\hline
\multirow{2}{5em}{2<z<3}& ZFOURGE &   $0.077 \pm0.0083$ &  $0.060\pm 0.010$ \\
                        & Kriek+07&                      & $0.308^{0.2286}_{0.147}$ \\
\hline
\multirow{2}{5em}{3<z<4}& ZFOURGE &   $0.0561 \pm 0.009$ &  $0.391 \pm 0.0698$ \\
                        & Marsan+16&                     &  $0.831 \pm 0.231$  \\
                        & This work&  $0.653 \pm 0.088$  &  $0.714\pm 0.149$   \\
\hline
\end{tabular}
\caption{ZFOURGE and SDSS AGNs were selected by X-ray+FIR+Radio,  \citep{Marsan2017ApJ} AGNs were selected by X-ray, [OIII] and H$\beta$ emission lines, Kriek+07 and our work AGNs were selected by [OIII] and H$\beta$ emission lines.}
\end{table}

Regarding the lower stellar mass range (10<log(M$_{\star}$/M$_{\sun})$<11), we can observe the $f_{\textrm{AGN}}$ remains similar through redshift, except for our recent AGN detections that boosted the $f_{\textrm{AGN}}=0.0561$ to $f_{\textrm{AGN}}=0.653 \pm 0.088$ between $3<z<4$. On the other hand, in the high stellar mass range (log(M$_{\star}$/M$_{\sun})$>11), the $f_{\textrm{AGN}}$ fluctuates, rising as the redshift increases, showing high $f_{\textrm{AGN}}$ in all the samples. Overall, the $f_{\textrm{AGN}}$ increase at high redshift for any stellar mass range, which is consistent with previous studies with smaller samples \citep{Marsan2017ApJ}.\\ 

We have considered the evident from line widths, luminosities and $[OIII]/H\beta$ ratio lying in the MEX-AGN region as pointing to all the galaxies having an origin of line emission from AGN. However not all galaxies meet all these criteria simultaneously. If we conservatively consider much stricter criteria considering AGNs with line ratios of [OIII]/H$_{\beta} > 10$ and a large velocity dispersion of  $\sigma_{5007} > 300$ km/s, we find an $f_{\textrm{AGN}} = 0.29 \pm 0.09$

We can calculate a robust lower limit on $f_{\textrm{AGN}}$ if we make the extreme assumption that all galaxies with photometric redshifts between $3<z<4$ (see Fig \ref{fig:AGNfschemma}) actually have spectroscopic redshifts in this range but lack emission lines and do not harbor AGN. This gives us an $f_{\textrm{AGN}}= 0.361 \pm 0.088$ and $f_{\textrm{AGN}}= 0.14 \pm 0.05 $ with the stricter criteria.

We can also calculate the f$_{AGN}$ for quiescent galaxies in our sample and consider quiescent galaxies from S18. Of the 22 galaxies in our sample, 3 were classified as quiescent galaxies in the UVJ plot (See Fig \ref{fig:UVJ}b). 
From S18 we have 12 quiescent galaxies, 2 of them in common with our sample, and one confirmed (UDS-8197) as a quiescent galaxy. Therefore we have 13 quiescent galaxies spectroscopically confirmed, where 3 of them host an AGN. This gives us an f$_{AGN}=0.23 \pm 0.11$ for quiescent galaxies in ZFOURGE for galaxies at $3<z<4$. This is significantly less than the f$_{AGN}$ for star-forming galaxies. We notice that the three AGNs in quiescent galaxies are within the stellar mass range of $10^{10}$<$M_{\odot}$<$10^{11}$, so the AGN fraction for these stellar mass range is $f_{\textrm{AGN}}=0.3 \pm 0.13$. 

\subsection{Shocks}

Finally, we briefly consider shocks as an explanation for the line ratios.
In \ref{sec:Emissionlined}, we applied the MEx line diagnosis method to determine the excitation sources of galaxies using ratios of strong optical emission lines of [OIII] and H$\beta$. AGNs lie at large [OIII]/H$\beta$ ratios on this diagram, where we found $64\%$ of our galaxy sample according to C15 criteria, or $95\%$  according to J14 (See fig \ref{fig:mass_exitation_colors}. Nevertheless, the dependence on metallicity of the line ratios diagnostic methods can drag the position of the AGN to lower masses in the MEx diagram \cite{Kewley_2013}.
 Separating between star formation, AGNs, and shocks is always difficult. Integral Field Spectroscopy is the most successful technique for separating these components among the various diagnostic diagrams proposed \citep{Allen_1998, Best200010.1046/j.1365-8711.2000.03027.x,refId0,Groves_2004, Feltre2016MNRAS.456.3354F,Jaskot2016ApJ...833..136J}. Nevertheless, we only measure [OIII] and H$\beta$ emission lines, so we cannot apply these diagnostics. Nevertheless, we calculate the bolometric luminosity of the galaxy sample (See fig \ref{fig:mass_exitation_colors} ) utilizing the [0III]5007 line, which provides a robust tracer of AGN power and is less contaminated by emission due to star formation activity. The high bolometric luminosity obtained for our sample makes it unlikely that the broad emission lines from our galaxies are powered only by shocks, as bolometric luminosities found for shocks in detailed studies with IFU data for the galaxy NGC 1068 are about L$_{bol}$=7-24 $\times 10^{41}$ erg/s \citep{dAgostino2019}.

\section{DISCUSSION}
\label{sec:DISCUSSION}
            
The high-AGN fraction --- 70$\%$ of our star-forming galaxies and 30\% of our quiescent galaxies --- shows that the period spanning $3<z<4$ seems to be a time of great activity for the growth of supermassive black holes for the massive galaxy population.

The observed AGN fractions exhibit an upward trend regardless of the stellar mass range. However, this trend is particularly remarkable for galaxies with log(M$_{\star}$/M$_{\sun})$>11. These findings align with similar studies with smaller samples of massive galaxies, where a f$_{AGN}$=0.8 (5/6) is found at $3<z<4$ by the strength of the line rations of their spectra, in combination with X-ray and radio detections \citep{Marsan2017ApJ}, highlighting the evolution of $f_{\textrm{AGN}}$ in cosmic time. 

The fraction of time that an AGN is active (duty cycle) in massive quiescent galaxies (log(M$_{\star}$/M$_{\sun})$>10) decreases at high stellar masses, while it increases with stellar mass for star-forming galaxies. At the same time, the fraction of galaxies with black holes accreting mass strongly evolves with redshift being comparable with star-forming galaxies at $z=2$ \citep{Aird_2017}. This could explain the different AGN fractions found in our sample regarding star-forming and quiescent galaxies, as short duty cycles make it harder to find AGNs in action at higher redshifts.

Hydrodynamical simulations can also give us constraints on the AGN population, although their calibration, sub-grid models of black holes and galaxy formation vary. \citep{Habouzit2021} compared the six Illustris, TNG100, TNG300, Horizon-AGN, EAGLE and SIMBA large-scale cosmological simulations \citep{Nelson_2017, Marinacci_2018, Springel_2017, Naiman_2018, Pillepich_2017, Kaviraj_2017, Schaye_2014, Dav__2019}. Finding a population of AGN that agrees with observational constraints at all redshifts and luminosities is challenging for simulations. Nevertheless, they found that the fraction of AGN is always higher at high redshifts for all the galaxy stellar mass bins. TNG and SIMBA simulations present similar trends to observations, although their AGN number densities peak at different redshifts. For simulated AGN with L$_{bol}> 10^{44}$ erg s$^{-1}$ in massive galaxies of log(M$_{\star}$/M$_{\sun})$>11 between $3<z<4$, TNG300 and SIMBA show an AGN fraction of about 80$\%$ \citep{Habouzit2021}. 

Although the triggering mechanism for the AGN activity still seems stochastic, AGNs are known to alter galaxies' star formation and be a driver of quenching for galaxies at low redshifts. The high AGN fraction found suggests it might be the dominant quenching mechanism at high redshifts \citep{Marchesini_2009, Aird_2017}.

There are two primary processes through which AGN feedback is thought to occur. One is called the "quasar-mode feedback" \citep{Silk1998}. In this process, wind ejects gases from galaxies within this mode, thereby impeding star formation. This process is widely believed to be prevalent in high-luminosity AGNs, such as those featuring QSOs, which are close to the Eddington limit. The other mode of feedback is known as "radio-mode feedback" or kinetic-mode feedback, as described by \citep{Fabian1996}. In this mode, low-luminosity AGNs, which make up less than one percent of the Eddington luminosity, heat the circumgalactic and halo gas through their radio jets, thereby preventing the gas from cooling. Compared to the quasar-mode feedback, radio-mode feedback is expected to keep the quiescence rather than reduce the star formation.

Low luminosity AGN have been found in X-ray and radio studies supporting the critical role of AGNs in star formation and quenching of massive galaxies \citep{Ito_2022}. The low range of $L_{\textrm{bol}}=10^{44}-10^{46}$ erg s$^{-1}$ in our sample of galaxies, would support the kinetic-mode feedback scenario as a quenching mechanism. Cosmological models and simulations require a feedback mechanism to replicate the properties of the massive galaxy population \citep{Springel2005, Croton2006, Somerville2008, Choi2015, Weinberger2018, Dav__2019}, as it is challenging to achieve the necessary energy levels through stellar feedback alone \citep{Bower2006}.

AGN feedback has been incorporated in various ways \citep{Somerville2015}, a common prescription in semianalytic models involves major mergers that trigger AGN-driven winds, expelling gas and eventually truncating star formation \citep{Hopkins2008}.
 Cosmological simulations such as IllustrisTNG suggest that AGN feedback is necessary as early as $z\sim6$ to produce massive quiescent galaxies at $z=4$ \citep{Hartley_2023, Qin2017}.

JWST has revealed an unexpected population of very red and compact sources with photometric redshifts $z>3$ \citep{Endsley_2023, Labbe2023, Furtak_2023,akins2023massive}, which spectroscopical follow-up have shown to contain moderate-luminosity heavily reddened AGN \citep{Harikane_2023, oesch2023jwst}. Also, JWST keeps finding AGNs at higher and higher redshifts, for instance 
at $z\sim5$ \citep{Onoue_2023, Kocevski_2023,Ubler2023};
in red galaxies at $z>5$ \citep{Greene2023}; 
low-luminosity quasars at  z$\sim$6 \citep{Ding_2022};
a triply-lensed red-quasar candidate at z$\sim$8 \citep{Furtak_2023};
H$\beta$ emission line in a galaxy at $z=8.7$ \citep{Larson_2023}. This shows that AGNs are more common than previously thought. 

The first census of  type-1 AGNs at $z>4$ \citep{harikane2023jwstnirspec} identified by JWST/NIRSpec deep spectroscopy indicates that about 5 $\%$ of the galaxies at $z=4-7$ harbor faint type-1 AGNs, while studies of local AGNs imply that only 1-2 $\%$ of galaxies with similar bolometric luminosities are type 1 AGNs. A high fraction of the broad-line AGNs indicates that the number density of such faint AGNs is higher than an extrapolation of the quasar luminosity function, implying a large population of AGNs, including type 1 and type 2, in the early universe. 

\section{CONCLUSIONS}
\label{sec:CONCLUSIONS}
We analyze a sample of 22 massive galaxies with masses $10<\log(M_{\star}/M_{\odot})<11$ at redshift $3<z<4$ from the ZFOURGE catalog. The galaxies' spectra show high velocity dispersions and large equivalent widths in their [OIII] and H$\beta$ emission lines.
We conducted tests to identify the sources of broad emission lines in our galaxy sample. We evaluated two scenarios of dynamical and AGN origin:

\begin{itemize}
    \item We find that a dynamical scenario is improbable based on the weak correlation between implied dynamical and stellar masses, i.e. that we are {\it not} seeing star-forming
    gas distributed around a galaxy-wide potential. 
    \item Five of our 22 galaxies were detected either in X-ray or radio in the ZFOURGE catalog, unequivocally indicating the presence of an AGN.
    \item We apply line diagnostic MEx diagrams to further distinguish between star-forming galaxies and AGNs. These show that in the majority of cases (19/22)  the sources are likely to be AGN. This is primarily  driven by the high [OIII]/H$\beta$ values. The other three objects lie in intermediate regions
    where the classification depends on the choice of MEx boundary but are still highly likely
    to be AGN.
    \item We derive bolometric luminosities from the [OIII] luminosities and these are in the range typical of AGN at lower redshifts.
    \item From the CIGALE SED fitting, we obtained a fraction of AGN IR luminosity to total IR luminosity higher than 0.2 for the entire sample, indicating AGN contribution in all the galaxies of the sample.
    \item The AGN fraction is high; overall we find fractions of 65.3$\pm 8.8 \% $ in the low mass bin increasing to 71.4 $\pm 14.9 \% $ in the high mass bin. This is higher than previous work and higher than at lower redshifts.
    \item Applying a stricter criteria to the AGNs considering line ratios of [OIII]/H$_{\beta} > 10$ and a large velocity dispersion of  $\sigma_{5007} > 300$ km/s, we find a lower limit of $f_{\textrm{AGN}} = 29 \pm 9 \% $.
    \item The lower limit for the AGN fraction is 36.1 $\pm 8 \%$, assuming that all galaxies with photometric redshift between $3<z<4$ have spectroscopic redshifts in this range but lack emission lines and do not harbour AGN and $f_{\textrm{AGN}}= 14 \pm 5 \% $ with stricter criteria.
    \item The AGN fraction for quiescent galaxies is considerably lower than the AGN for star-forming galaxies, with a value of 23 $\pm 11 \% $ for galaxies with stellar mass log$(M_{\star}/M_{\odot})>11$ and 30 $\pm 13 \%$ for galaxies with $10<\log(M_{\star}/M_{\odot})<11$.

\end{itemize}

Overall our results point to a ubiquity of AGN in massive galaxies at $3<z<4$ and is consistent with a
picture in which the massive quenched galaxies at this epoch would have been produced by AGN quenching.
Future work could improve this. For our sample it is now possible to access H$\alpha$, [NII] and [SII]
lines using the NIRSPEC spectrograph on the James Webb Space Telescope thus accessing the full 
BPT diagram. Multiple emission lines can be measured from rest-frame UV to near-infrared which will allow
improved diagnostics of the ionizing mechanisms, the density and pressure of gas in the narrow line region,
and its metallicity. We predict that the emission lines in our sample
would arise from point sources at the center of each galaxy; this could be tested using the NIRSPEC Integral
Field Spectrograph or with ground based IFS. The latter could also measure galaxy-wide kinematics and hence derive scaling 
relations between true dynamical mass and central black hole mass.

\begin{landscape}
\begin{table}
	\centering
	\caption{Final properties of the galaxy sample.}
	\label{tab:FINALTABLE}
	\begin{tabular}{llllllllllll} 
\hline		
CATAID	     & $\log(M_{\star}/M_{\odot})$ & $\sigma_{5007}$ & $\log$(SFR)  & $r_{e}$	 &   n  & $\log(M_{dyn}/M_{\odot})$  & $\log($[OIII]/H$\beta$) & [OIII]$_{EW}$   & $H\beta_{EW}$ & $L_{bol}$ & $f_{AGN}^{IR}$  \\

 	         &                             & (km $s^{-1}$)   & ($M_{\odot}/yr$)&(kpc)	 &      &       &                          &   (\AA)         &  (\AA)  & erg s$^{-1}$ &\\
\hline	
\hline	
COSMOS-2561	 & $11.06 \pm 0.02$  &	$277 \pm 54$ &$1.1$ & $3.0 \pm 0.6 $  & 2.6 & $11.5 \pm 0.3$ & $0.71 \pm 0.15$ & $161 \pm 1$   & $60 \pm 13$  & 1.3 $\times 10^{45}$  & $0.49 \pm 0.1$\\
COSMOS-4214	 & $10.72 \pm 0.1$   &	$169 \pm 39$ &$1.6$ & $2.6 \pm 0.4 $  & 1.6 & $11.1 \pm 0.2$ & $0.35 \pm 0.04$ & $964 \pm 206$ &$141.2\pm5$   & 1.4 $\times 10^{45}$  & $0.9  \pm 0.0001$\\
COSMOS-5874	 & $10.85 \pm 0.02$  &	$195 \pm 65$ &$2.1$ & $7.2 \pm 1.8 $  & 1.7 & $11.6 \pm 0.5$ & $1.79 \pm 3.03 $& $47  \pm 2 $  &$89\pm13$     & 9.8 $\times 10^{44}$  & $0.9  \pm 0.0002$\\
COSMOS-12173 & $11.33 \pm 0.02$  &	$309 \pm 37$ &$1.9$ & $1.8 \pm 0.0 $  & 2.9 & $11.4 \pm 0.0$ & $0.97 \pm 0.06$ & $661 \pm 91 $ &$45\pm2$      & 8.4 $\times 10^{45}$  & $0.85 \pm 0.049$\\
COSMOS-20133 & $10.71 \pm 0.02$  &	$60  \pm 34$ &$0.5$ & $0.2 \pm 0.0 $  & 3.0 & $9.07 \pm 0.1$ & $0.80 \pm 0.02$ & $718 \pm 12 $ &$98\pm12$     & 2.1 $\times 10^{45}$  & $0.803\pm 0.018$\\ 
COSMOS-20169 & $10.82 \pm 0.02$  &  $214 \pm 39$ &$1.1$ & $4.1 \pm 1.7 $  & 2.1 & $11.4 \pm 0.5$ & $0.46 \pm 0.17$ & $214 \pm 79 $ &$10\pm5$      & 6.2 $\times 10^{44}$   & $0.9  \pm  0.0001$\\
CDFS-6691    & $11.01 \pm 0.02$  &  $401 \pm 34$ &$2.3$ & $12.9\pm 3.5 $  & 8.0 & $12.2 \pm 0.0$ & $1.87 \pm 0.42$ &  $82 \pm 5$   &$50 \pm 5$    & 3.7 $\times 10^{45}$  & $0.9  \pm 0.0001$\\  
CDFS-13954   & $11.29 \pm 0.02$  &  $442 \pm 37$ & $2.1$& $1.3 \pm 0.0 $  & 3.7 & $11.5 \pm 0.0$ & $0.99 \pm 0.27$ &  $153\pm 13$  &$14 \pm3$     & 4 $\times 10^{45}$    & $0.9  \pm 0.0004$\\
CDFS-7535	 & $10.25 \pm 0.24$  &	$330 \pm 37$ &$2.9$ & $8.8 \pm 2.1 $  & 10.1& $10.2 \pm 0.2$ & $0.68 \pm 0.12$ & $290 \pm 49$  &$73\pm16$     &1.6 $\times 10^{46}$   & $0.9  \pm 0.00001$\\
CDFS-7752 	 & $10.78 \pm 0.03$  &	$84  \pm 34$ &$2.7$ & $1.4 \pm 0.0 $  & 2.9 & $10.2 \pm 0.1$ & $0.39 \pm 0.01$ & $214 \pm 2 $  &$59\pm4$      &3.7 $\times 10^{45}$   & $0.89 \pm 0.028$\\
CDFS-8428	 & $11.44 \pm 0.02$  &	$352 \pm 45$ &$2.3$ & $0.7 \pm 0.0 $  & 6.6 & $10.9 \pm 0.1$ & $0.87 \pm 0.05 $& $768 \pm 15$  &$46\pm13$     &1 $\times 10^{46}$     & $0.9  \pm 0.0005$\\
CDFS-9332	 & $11.0  \pm 0.06$  &	$186 \pm 37$ &$2.1$ & $0.5 \pm 0.0 $  & 5.7 & $11.0 \pm 0.1$ & $1.59 \pm 0.47$ & $523 \pm 11$  &$77\pm14$     &1 $\times 10^{46}$     & $0.6  \pm 0.1 $\\
CDFS-9833	 & $10.09 \pm 0.08$  &  $105 \pm 34$ &$2.4$ & $2.1 \pm 0.0 $  & 3.0 & $10.5 \pm 0.0$ & $0.81 \pm 0.02$ & $2716\pm 98$  &$303\pm16$    &2.1$ \times 10^{46}$   & $0.89 \pm 0.02 $\\
UDS-3417	 & $11.3  \pm 0.04$  &	$325 \pm 37$ &$1.9$ & $2.2 \pm 0.2 $  & 7.5 & $11.3 \pm 0.1$ & $0.92 \pm 0.02$ & $1050\pm 34$  &$15\pm34$     &1.2$ \times 10^{46}$   & $0.31 \pm 0.2 $\\
UDS-6023	 & $11.08 \pm 0.03$  &	$90  \pm 37$ &$1.0$ & $0.3 \pm 0.0 $  & 7.6 & $9.35 \pm 0.2$ & $0.80 \pm 0.10$ & $417 \pm 1$   &$48\pm18$     &1.3$ \times 10^{45}$   & $0.89 \pm 0.07 $\\
UDS-6226	 & $10.22 \pm 0.33$  &	$428 \pm 36$ &$1.8$ & $8.7 \pm 1.6 $  & 3.4 & $12.3 \pm 0.1$ & $0.60 \pm 0.38$ & $110 \pm 33$  &$122\pm46$    &1.5$ \times 10^{45}$   & $0.7  \pm 0.133 $\\
UDS-6639     & $11.37 \pm 0.02$  &  $214 \pm 59$ &$2.6$ & $7.46\pm 0   $  & 2.8 & $11.7 \pm 0.0$ & $0.36 \pm 0.86$ & $192.87 \pm 91$& $167.3 \pm 91$&5$ \times 10^{44}$   & $0.68 \pm 0.28 $\\
UDS-8092     & $11.4  \pm 0.05$  &  $396 \pm 51$ &$1.2$ & $1.5 \pm 0   $  & 1.3 & $11.6 \pm 0.0$ & $1.46 \pm 0.03$ & $116 \pm 9$   & $9 \pm 1$      &1.2 $\times 10^{45}$ & $0.71 \pm 0.07 $\\
UDS-8197     & $10.93 \pm 0.02$  &  $395 \pm 34$ &$2.3$ & $0.7 \pm 0.2 $  & 8   & $11.1 \pm 0.4$ & $ 1.18\pm 0.4$  & $513 \pm 45$  & $31 \pm 4.5$   &2.4 $\times 10^{45}$ & $0.89 \pm 0.01 $\\
UDS-8844	 & $10.99 \pm 0.03$  &	$138 \pm 41$ &$2.2$ & $2.5 \pm 0.3 $  & 2.1 & $10.9 \pm 0.2$ & $-0.09\pm 0.22$ & $16  \pm 9$   &$25\pm9$        &5 $\times 10^{45}$   & $0.78 \pm 0.05 $\\
UDS-10644	 & $10.96 \pm 0.02$  &	$272 \pm 52$ &$1.5$ & $3.1 \pm 1.3 $  & 3.6 & $11.5 \pm 0.5$ & $0.55 \pm 0.36$ & $544 \pm 54$  &$46\pm34$       &1.7 $\times 10^{45}$ & $0.89 \pm 0.001$\\
UDS-14075	 & $9.26  \pm 0.13$  &	$110 \pm 38$ &$1.3$ & $0.9 \pm 0.1 $  & 1.0 & $10.3 \pm 0.2$ & $0.24 \pm 0.29$ & $364 \pm 82$  &$153\pm95$      &7.4$\times 10^{44}$  & $0.25 \pm 0.26 $\\

\hline
\end{tabular}
\end{table}
\end{landscape}

\section*{Acknowledgements}

We acknowledge financial support from NASA through grant JWST-ERS-1324. T.N., K. G., and C.J. acknowledge support from the Australian Research Council Laureate Fellowship FL180100060. 

\section{Data availability}

The data underlying this article will be shared on reasonable request to the corresponding author.

\bibliographystyle{mnras}
\bibliography{MGAGN} 





\appendix
\section{Apendix}
\label{sec:Apendix}

\begin{figure*}
	\includegraphics[width=\textwidth]{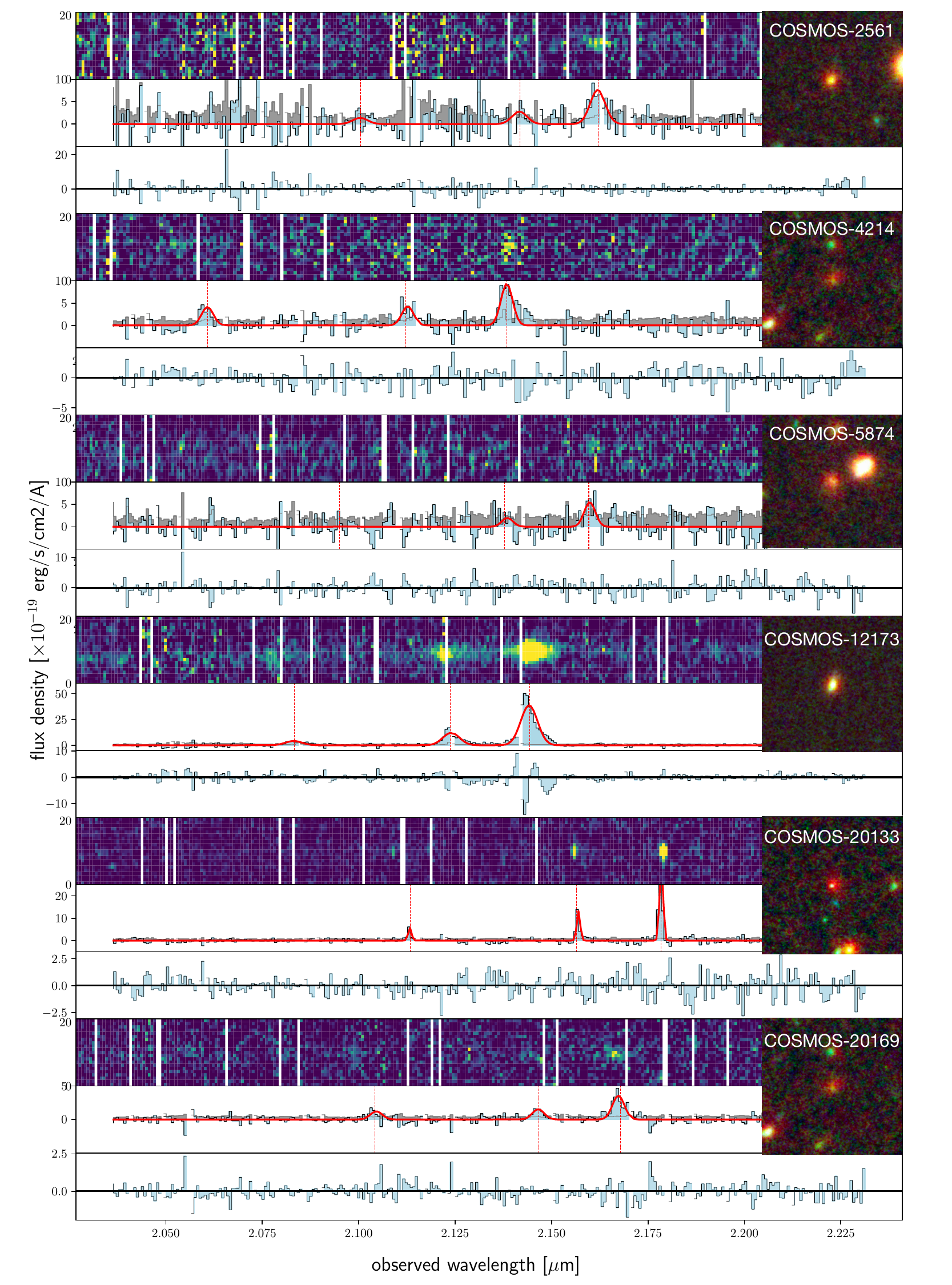}
    \caption{Best fit of the H$\beta$, [OIII]5007 and [OIII]4959 emission lines for each galaxy of COSMOS field. For each example, we show on the top left panel the 2D spectrum from keck/MOSFIRE, on the middle panel the spectrum with the best fit in red, on the bottom panel the residuals, and in the right corner a WFC3/F160W HST image stamp.}
    \label{fig:spectra_figure_cosmos}
\end{figure*}

\begin{figure*}
	\includegraphics[width=\textwidth]{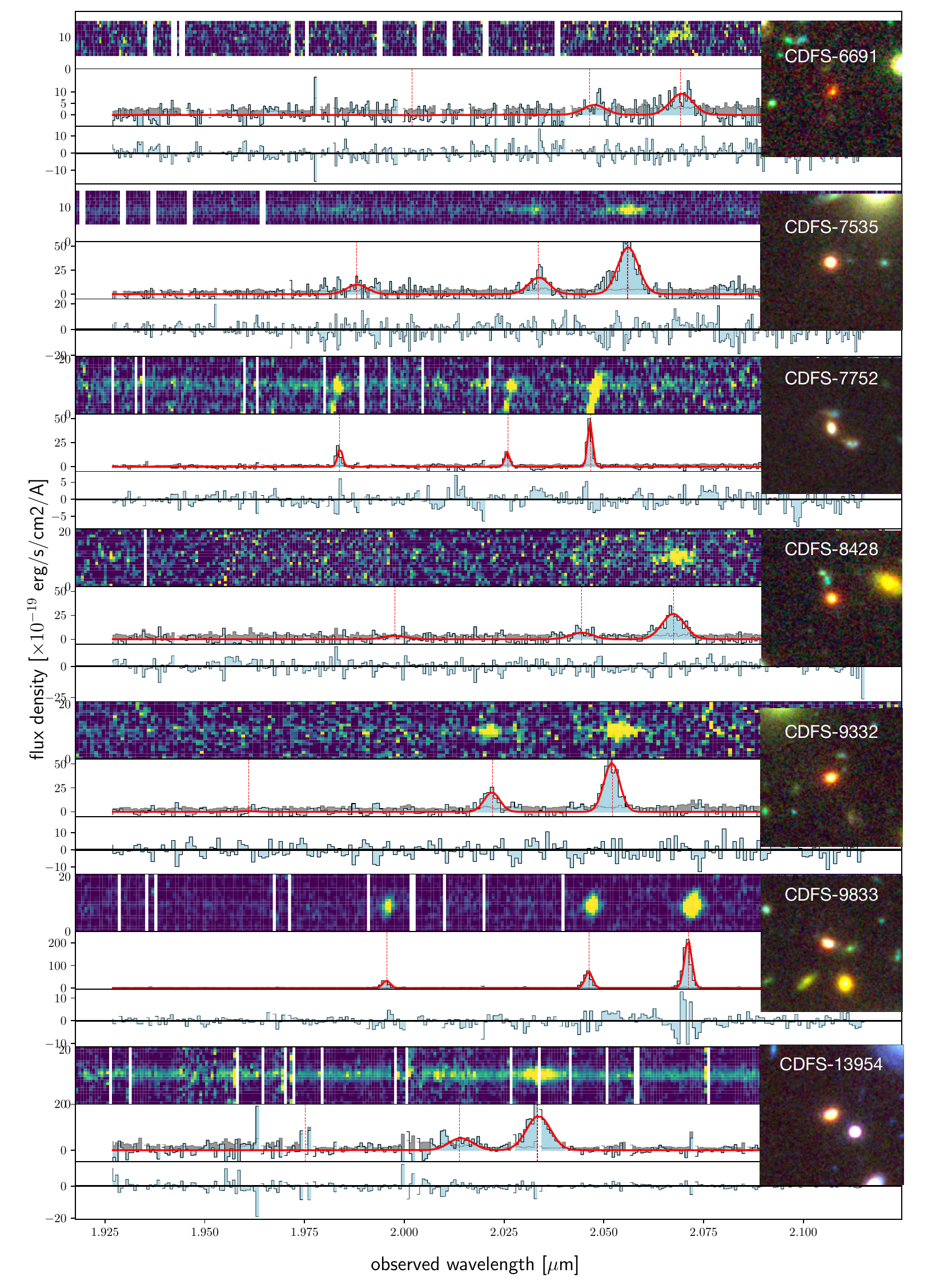}
    \caption{Best fit of the H$\beta$, [OIII]5007 and [OIII]4959 emission lines for each galaxy of CDFS field. For each example, we show on the top left panel the 2D spectrum from keck/MOSFIRE, KMOS for CDFS-6691 and CDFS-7535 on the middle panel the spectrum with the best fit in red, and on the bottom panel the residuals. In the right corner, we show the WFC3/F160W HST image stamp.}
    \label{fig:spectra_figure_cdfs}
\end{figure*}

\begin{figure*}
	\includegraphics[width=\textwidth]{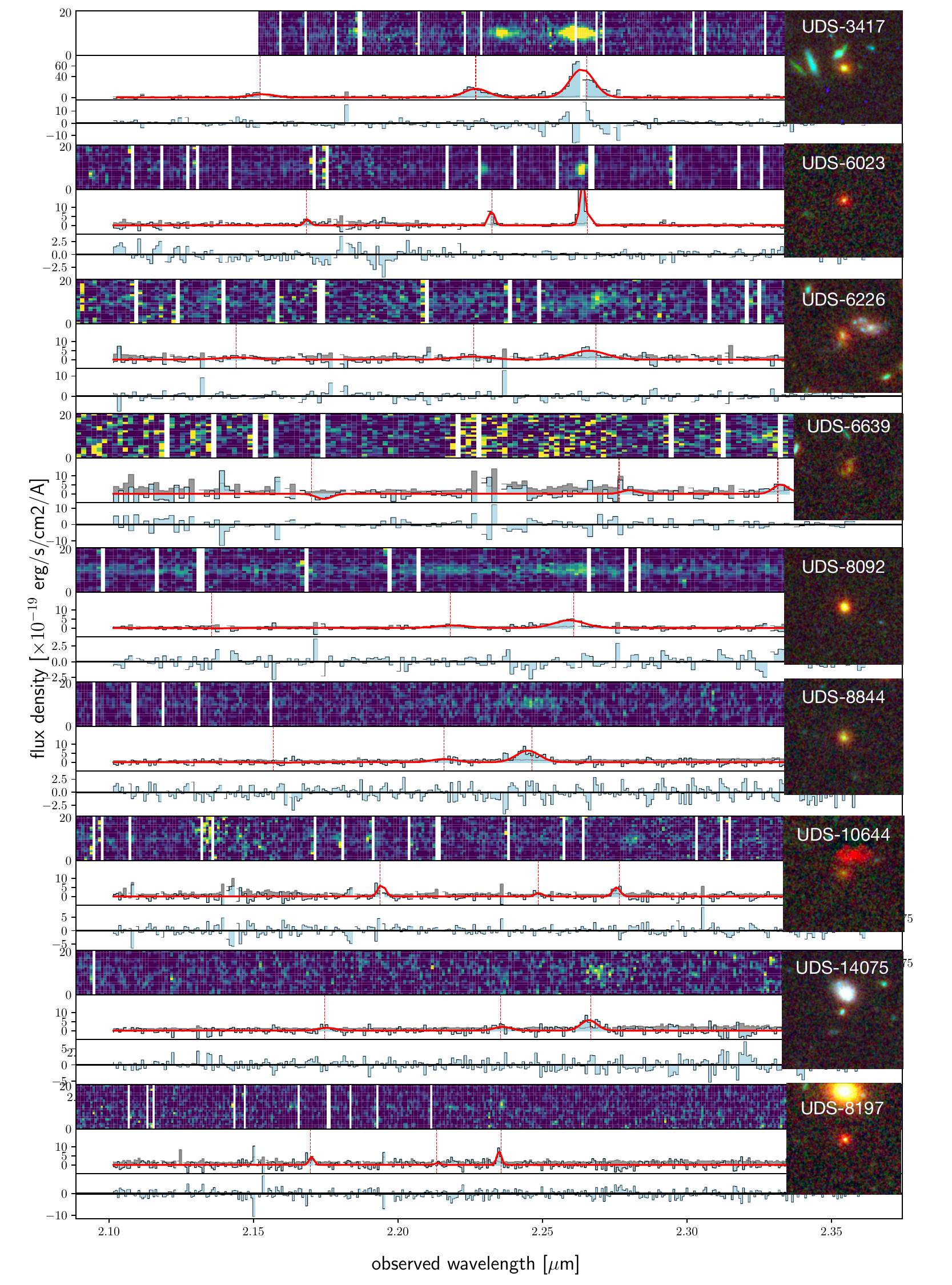}
    \caption{Best fit of the H$\beta$, [OIII]5007 and [OIII]4959 emission lines for each galaxy of UDS field. For each example, we show on the top left panel the 2D spectrum from keck/MOSFIRE, on the middle panel the spectrum with the best fit in red, and on the bottom panel the residuals. In the right corner, we show the WFC3/F160W HST image stamp.}
    \label{fig:spectra_figure_uds}
\end{figure*}

\bsp	
\label{lastpage}
\end{document}